\documentclass{emulateapj}
\usepackage{natbib}
\usepackage{epsfig}
\usepackage{graphicx}
\usepackage{subfigure}
\usepackage{float}
\usepackage{amsmath}
\usepackage{color}
\usepackage{amssymb}
\usepackage{amsfonts}
\usepackage{units}
\usepackage[colorlinks,linkcolor=blue,anchorcolor=green,citecolor=blue]{hyperref}
\shorttitle{Fast Radio Burst Cosmology}
\shortauthors{Yang \& Zhang}
\begin{document}

\title{Extracting host galaxy dispersion measure and constraining cosmological parameters using fast radio burst data}

\author{Yuan-Pei Yang\altaffilmark{1} and Bing Zhang\altaffilmark{1,2,3}}

\affil{$^1$Kavli Institute of Astronomy and Astrophysics, Peking University, Beijing 100871, China; yypspore@gmail.com;\\
$^2$ Department of Astronomy, School of Physics, Peking University, Beijing 100871, China \\
$^3$ Department of Physics and Astronomy, University of Nevada, Las Vegas, NV 89154, USA; zhang@physics.unlv.edu}

\begin{abstract}
The excessive dispersion measures (DMs) and high Galactic latitudes of fast radio bursts (FRBs) hint toward a cosmological origin of these mysterious transients. Methods of using measured DM and redshift $z$ to study cosmology have been proposed, but one needs to assume a certain amount of DM contribution from the host galaxy (DM$_{\rm HG}$) in order to apply those methods. We introduce a slope parameter $\beta(z) \equiv d \ln \left< {\rm DM}_{\rm E} \right> / d \ln z$ (where DM$_{\rm E}$ is the observed DM subtracting the Galactic contribution), which can be directly measured when a sample of FRBs have $z$ measured. We show that $\left< {\rm DM_{HG}}\right>$ can be roughly inferred from $\beta$ and the mean values, $\overline{\rm \left<DM_{\rm E}\right>}$ and $\bar z$, of the sample. Through Monte Carlo simulations, we show that the mean value of local host galaxy DM, $\left<\rm{DM_{HG,loc}}\right>$, along with other cosmological parameters (mass density $\Omega_m$ in the $\Lambda$CDM model, and the IGM portion of the baryon energy density $\Omega_b f_{\rm IGM}$) can be independently measured through  MCMC fitting to the data. 
\end{abstract}


\section{Introduction}

Fast radio bursts (FRBs) are a new mysterious class of radio transients observed at frequencies around $1~\unit{GHz}$. They are characterized by short intrinsic durations ($\sim 1~\unit{ms}$), 
large dispersion measures (${\rm DM}\gtrsim 500~\unit{pc~cm^{-3}}$), and high Galactic latitudes \citep{lor07,kea12,tho13,bur14,spi14,spi16,mas15,pet15,rav15,cha16,kea16}.
The observed DMs have a large excess with respect to the Galactic value in the high Galactic latitude directions from which the FRBs are observed, suggesting an extragalactic or even a cosmological origin \citep{tho13,kul14}. The observed DM should have a large contribution from the intergalactic medium (IGM). If redshifts of FRBs can be measured, one may combine the DM and $z$ information to perform cosmological studies \citep{den14,gao14,zho14,zhe14}.\footnote{So far, only one FRB has its host galaxy's redshift $z=0.492$ claimed \citep{kea16}. The case is however controversial \citep{wil16,ved16,li16}.} 

In previous works \citep{den14,zho14,gao14}, in order to constrain the cosmological parameters with FRB observations, one needs to first subtract the host galaxy contribution, ${\rm DM_{HG}}$  (which includes contributions from the host galaxy interstellar medium and the plasma associated with the FRB source), from the observed value, DM$_{\rm obs}$, in order to 
obtain the DM from the intergalactic medium, ${\rm DM_{IGM}}$. If one has ${\rm DM_{IGM}}$ and redshift $z$ measured for a sample of FRBs, many interesting cosmological applications are possible. However, ${\rm DM_{HG}}$ is a poorly known parameter, which depends on the type of the host galaxy, the site of FRB in the host galaxy, the inclination angle of the galaxy disk, and the near-source plasma contribution \citep{gao14,xu15}. Another complication is ${\rm DM_{IGM}}$ depends on  $\Omega_b f_{\rm{IGM}}$ \citep{den14,gao14,zho14}, where $\Omega_b$ is the current baryon mass density fraction of the universe and $f_{\rm{IGM}}$ is the fraction of baryon mass in the intergalactic medium. Both values have to be inferred from other cosmological observations. 

In this paper, we study the first derivative of the ${\rm DM}-z$ relation, and find that the $\log {\rm DM_E}-\log z$ slope (where ${\rm DM_E}$ is the extragalactic dispersion measure of the FRB), $\beta\equiv d\ln\left<{\rm DM_E}\right>/d\ln z$, can be used to infer $\left<{\rm DM_{HG}}\right>$. We further show that $\left<{\rm DM_{HG,loc}}\right>$ and cosmological parameters ($\Omega_m$ in $\Lambda$CDM cosmology and $\Omega_b f_{\rm IGM}$) can be independently inferred by applying a Markov Chain Monte Carlo (MCMC) fit to a sample of FRBs whose DM and $z$ are measured.

\section{Method}

The observed dispersion measure of an FRB is given by \citep{den14,gao14}
\begin{eqnarray}
{\rm DM_{obs}}={\rm DM_{MW}}+{\rm DM_{IGM}}+{\rm DM_{HG}},
\end{eqnarray}
where ${\rm DM_{MW}}$, ${\rm DM_{IGM}}$ and ${\rm DM_{HG}}$ denote the contributions from the Milk Way, intergalactic medium and the FRB host galaxy (including interstellar medium of the host and the near-source plasma), respectively. ${\rm DM_{MW}}$ can be well constrained with the Galactic pulsar data \citep{tay93}, and is a strong function of the Galactic latitude $|b|$, e.g. ${\rm DM_{MW}}\sim 1000~\unit{pc~cm^{-3}}$ for $|b|\sim 0^\circ$, and ${\rm DM_{MW}}< 100~\unit{pc~cm^{-3}}$ for $|b|> 10^\circ$. For a well-localized FRB, ${\rm DM_{MW}}$ can be extracted with reasonable certainty. 
We then define extragalactic dispersion measure of an FRB as
\begin{eqnarray}
{\rm DM_E}\equiv{\rm DM_{obs}}-{\rm DM_{MW}}={\rm DM_{IGM}}+{\rm DM_{HG}}.\label{dme}
\end{eqnarray}

Since ${\rm \Lambda CMD}$ is consistent with essentially all observational constraints, in the rest of the paper we focus on this model with $\Omega_m+\Omega_\Lambda=1$ enforced.\footnote{For more complicated dark energy models, the method proposed in this paper may be also employed, but additional simulations are needed to see how well different dark energy models may be constrained.}
Considering local inhomogeneity of IGM, we define the mean dispersion measure of the IGM, which is given by \citep{den14}
\begin{eqnarray}
\langle{\rm{DM_{IGM}}}\rangle=K_{\rm IGM} \int_0^z\frac{f_e(z^\prime)(1+z^\prime)}{\sqrt{\Omega_m(1+z^\prime)^3+\Omega_\Lambda}}dz^\prime,
\label{dmigm}
\end{eqnarray}
where
\begin{equation}
K_{\rm IGM} \equiv \frac{3cH_0\Omega_bf_{\rm{IGM}}}{8\pi Gm_p},
\label{eq:K}
\end{equation}
$H_0$ is the current Hubble constant, $\Omega_b$ is the current baryon mass density fraction of the universe, $f_{\rm{IGM}}$ is the fraction of baryon mass in the intergalactic medium, $f_e(z)=(3/4)y_1\chi_{e,\rm{H}}(z)+(1/8)y_2\chi_{e,\rm{He}}(z)$, $y_1\sim 1$ and $y_2\simeq 4-3y_1\sim 1$ are the hydrogen and helium mass fractions normalized to 3/4 and 1/4, respectively, and $\chi_{e,\rm{H}}(z)$ and $\chi_{e,\rm{He}}(z)$ are the ionization fractions for hydrogen and helium, respectively. For FRBs at $z<3$, both hydrogen and helium are fully ionized \citep{mei09,bec11}. One then has $\chi_{e,\rm{H}}(z)=\chi_{e,\rm{He}}(z)=1$, and $f_e(z)\simeq 7/8$. 

\begin{figure}[H]
\centering
	\begin{subfigure}[]	
	\centering
	\includegraphics[angle=0,scale=0.9]{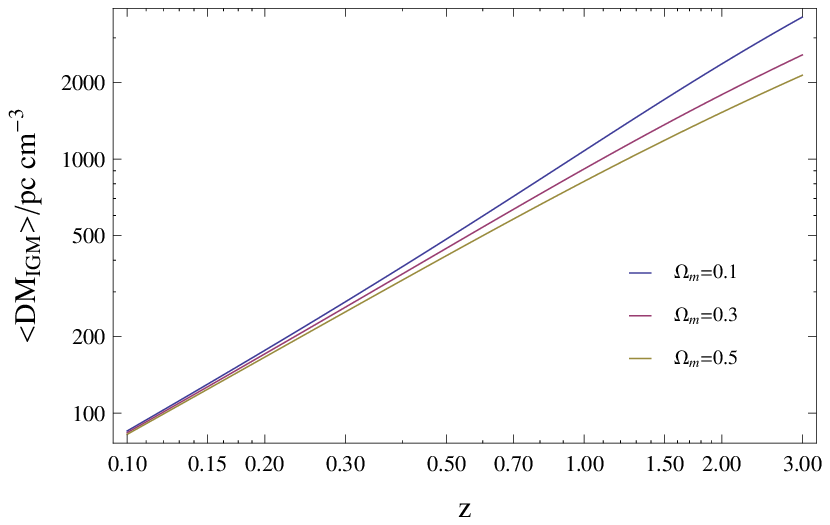}
	\end{subfigure}
	\begin{subfigure}[]
	\centering
	\includegraphics[angle=0,scale=0.9]{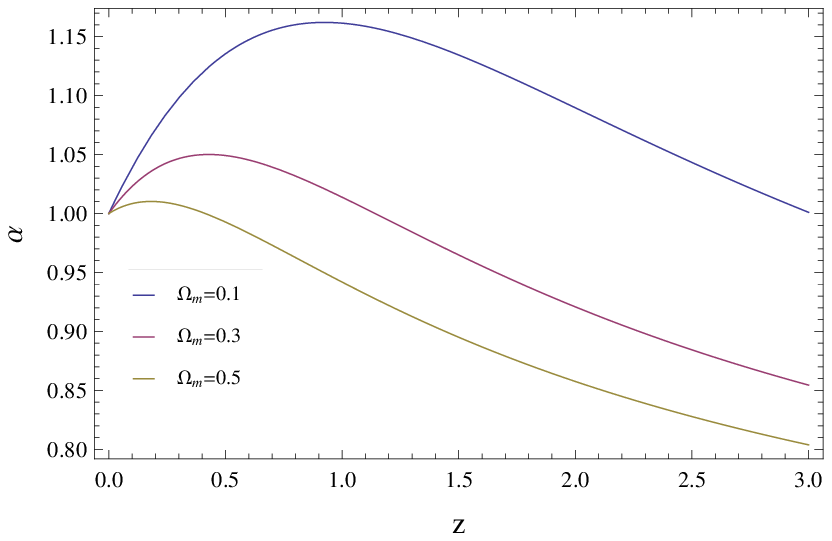}
	\end{subfigure}
\caption{(a). $\langle{\rm DM_{IGM}}\rangle$-$z$ relation. We adopted the best-constrained values of the following parameters \citep{pla15}:  $H_0=67.7~\unit{km~s^{-1}Mpc^{-1}},~\Omega_b=0.049,~f_{\rm{IGM}}=0.83$. (b). $\alpha$-$z$ relation. The blue, red and yellow lines denote $\Omega_m=0.1,\,0.3,\,0.5$, respectively. }\label{fig1}
\end{figure}

In an effort of investigating the first derivative of the ${\rm DM}-z$ relation, we first define
\begin{eqnarray}
\alpha(z)&\equiv&\frac{d\ln \langle\rm{DM_{IGM}}\rangle}{d\ln z}\nonumber\\
&=&\frac{zf_e(z)(1+z)/\sqrt{\Omega_m(1+z)^3+\Omega_\Lambda}}{\int_0^zf_e(z')(1+z')/\sqrt{\Omega_m(1+z')^3+\Omega_\Lambda}dz'}.\label{alpha1}
\end{eqnarray}
Since $f_e(z)\simeq 7/8$ for $z<3$, $\alpha$ essentially depends only on the cosmological parameters $(\Omega_m, \Omega_\Lambda)$. The $\langle{\rm DM_{IGM}}\rangle-z$ relation and $\alpha$ as a function of $z$ are presented in Fig.\ref{fig1} for $\Omega_m = 0.1, 0.3, 0.5$, respectively. One can see that $\alpha$ is around 1 at $z\lesssim1$. It initially rises and monotonically decreases with $z$ after reaching a peak. 

\begin{figure}[H]
\centering
	\begin{subfigure}[]
	\centering
	\includegraphics[angle=0,scale=0.9]{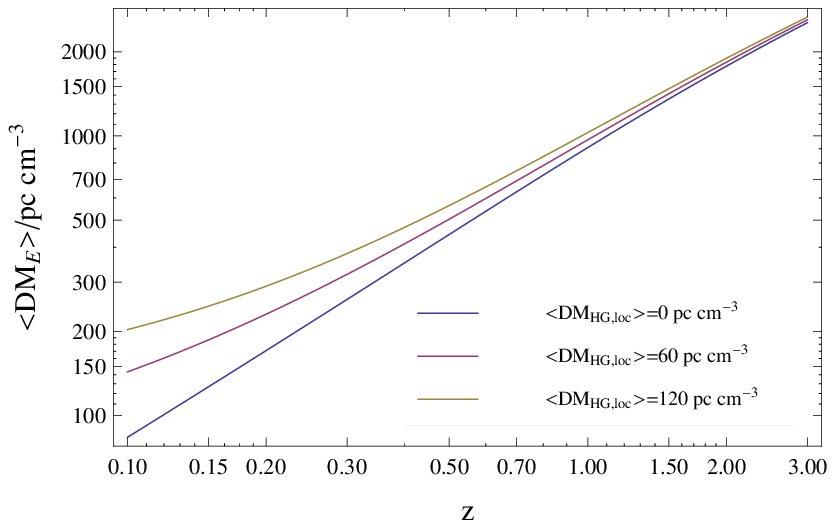}
	\end{subfigure}
	\begin{subfigure}[]
	\centering
	\includegraphics[angle=0,scale=0.9]{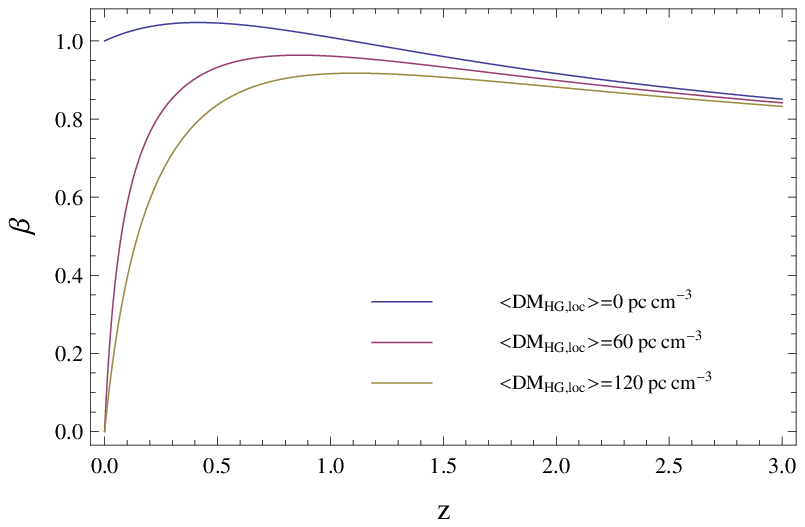}
	\end{subfigure}
\caption{(a). $\langle{\rm DM_{E}}\rangle$-$z$ relation. We adopt the same parameters as Fig.\ref{fig1} and $\Omega_m=0.31$. (b). $\beta$-$z$ relation. The blue, red and yellow lines denote $\langle{\rm DM_{HG,loc}}\rangle=0,\,60,\,120~\unit{pc~cm^{-3}}$, respectively.}\label{fig2}
\end{figure}

Observationally, one cannot directly measure  ${\rm DM_{IGM}}$, so that $\alpha$ cannot be directly measured. Since ${\rm DM_{E}}$ and $z$ are the directly measured parameters, we next define
\begin{eqnarray}
\beta(z) & \equiv & \frac{d\ln \langle\rm{DM_E}\rangle}{d\ln z} \nonumber \\
& = & \frac{z}{\langle\rm{DM_E}\rangle}\left(\frac{d\langle\rm{DM_{IGM}}\rangle}{dz}+\frac{d\langle\rm{DM_{HG}}\rangle}{dz}\right). 
\end{eqnarray}
In view of the dispersion of both $\rm{DM_E}$ and $\rm{DM_{HG}}$ in different directions at a same $z$, we have introduced the average values $\langle\rm{DM_E}\rangle$ and $\langle\rm{DM_{HG}}\rangle$ at redshift $z$ (in practice they are the average values in a certain redshift bin centered around $z$). 
For a host galaxy at redshift $z$, due to cosmological redshift and time dilation, its observed ${\rm DM_{HG}}$ is a factor of $1/(1+z)$ of the local one $\rm{DM_{HG,loc}}$ \citep{iok03,den14}. If we assume that the properties of FRB host galaxies have no significant evolution with redshift, then $d\langle{\rm DM_{HG}}\rangle/dz\simeq-\langle{\rm DM_{HG,loc}}\rangle/(1+z)^2$, and
\begin{eqnarray}
\beta(z)&=&\frac{\langle{\rm DM_E}\rangle-\langle{\rm DM_{HG,loc}}\rangle/(1+z)}{\langle{\rm DM_E}\rangle}\alpha(z)
\nonumber \\
& - &\frac{\langle\rm{DM_{HG,loc}}\rangle}{\langle\rm{DM_{E}}\rangle}\frac{z}{(1+z)^2}.
\label{beta}
\end{eqnarray}
One can see that due to the non-zero value of $\langle{\rm DM_{HG,loc}}\rangle$ and a $z$-dependent $\langle{\rm DM_E}\rangle$, 
$\beta(z)$ shows a different behavior from $\alpha(z)$ (Fig.\ref{fig2}): $\beta(z)\sim0$ for $z\ll1$, and $\beta(z)\sim\alpha(z)$ for $z\gg1$.

Since for standard cosmological parameters, $\alpha\simeq 1$ at $z\lesssim1$, one can estimate $\langle{\rm DM_{HG,loc}}\rangle$ using a sample of FRBs at low redshifts. Let us consider a sample of FRBs with $z<z_c\simeq0.5$. 
According to Eq.(\ref{beta}), one can derive
\begin{eqnarray}
\langle{\rm DM_{HG,loc}}\rangle\simeq\frac{(1-\bar\beta)(1+\overline{z})^2}{1+2\overline{z}}\overline{\langle{\rm DM_{E}}\rangle}.\label{dmhgp}
\end{eqnarray}
where the over-line symbols denote an average over all the FRBs in the sample at $z<z_c$, and $\bar\beta$ is the slope of linear fitting in the $z$ range in log-log space. In particular, for $z\ll1$, one has
\begin{eqnarray}
\langle{\rm DM_{HG,loc}}\rangle\simeq(1-\bar\beta)\overline{\langle{\rm DM_{E}}\rangle}.
\end{eqnarray}
One can see that a sample of FRBs at low $z$ would give a rough estimate of the host galaxy DM, $\langle{\rm DM_{HG,loc}}\rangle$.

On the other hand, due to $\langle{\rm DM_E}\rangle\gg\langle{\rm DM_{HG,loc}}\rangle$ at $z\gtrsim1$, one has $\alpha(z)\simeq\beta(z)$, which means that one can obtain the cosmological parameters by measuring $\beta$ at high redshift. In particular, for flat $\Lambda$CDM models, $\beta$ at high-$z$ would give a direct measure of $\Omega_m$. Finally, the absolute value of $\langle{\rm DM_E}\rangle$ at a given $z$ depends on the $K_{\rm IGM}$ parameter (Eq.(\ref{eq:K})). {\em As a result, the three unknown parameters, ${\rm DM_{HG,loc}}$, $\Omega_m$, and $K_{\rm IGM}$, are defined by different properties of the $\log {\rm DM_E}-\log z$ plot, and therefore can be independently inferred from the $(\langle{\rm DM_E}\rangle, z)$ data of a sample of FRBs.}

\section{Monte Carlo simulations}

\begin{figure}[H]
\centering
	\begin{subfigure}[]
	\centering
	\includegraphics[angle=0,scale=0.3]{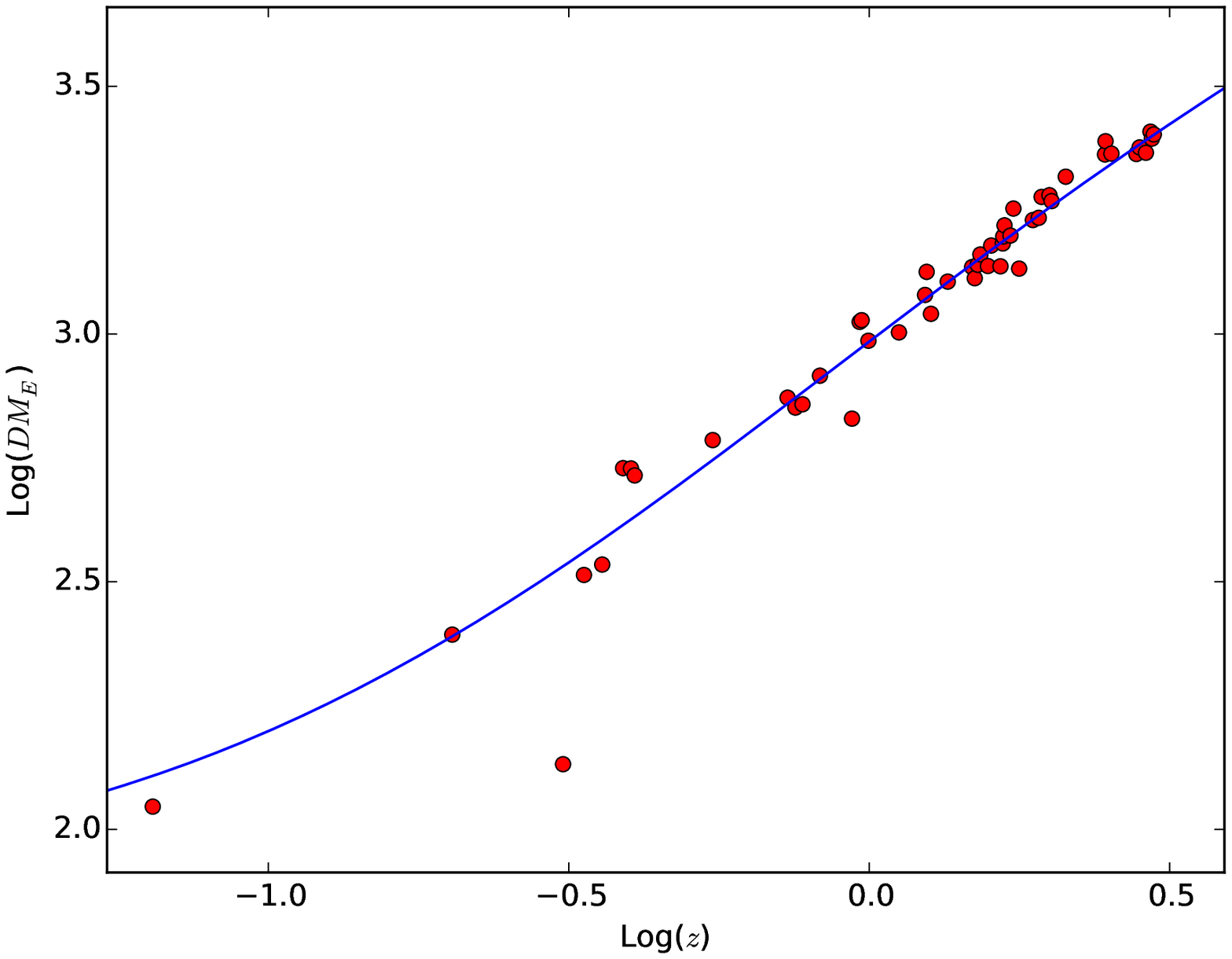}
	\end{subfigure}
	\begin{subfigure}[]
	\centering
	\includegraphics[angle=0,scale=0.3]{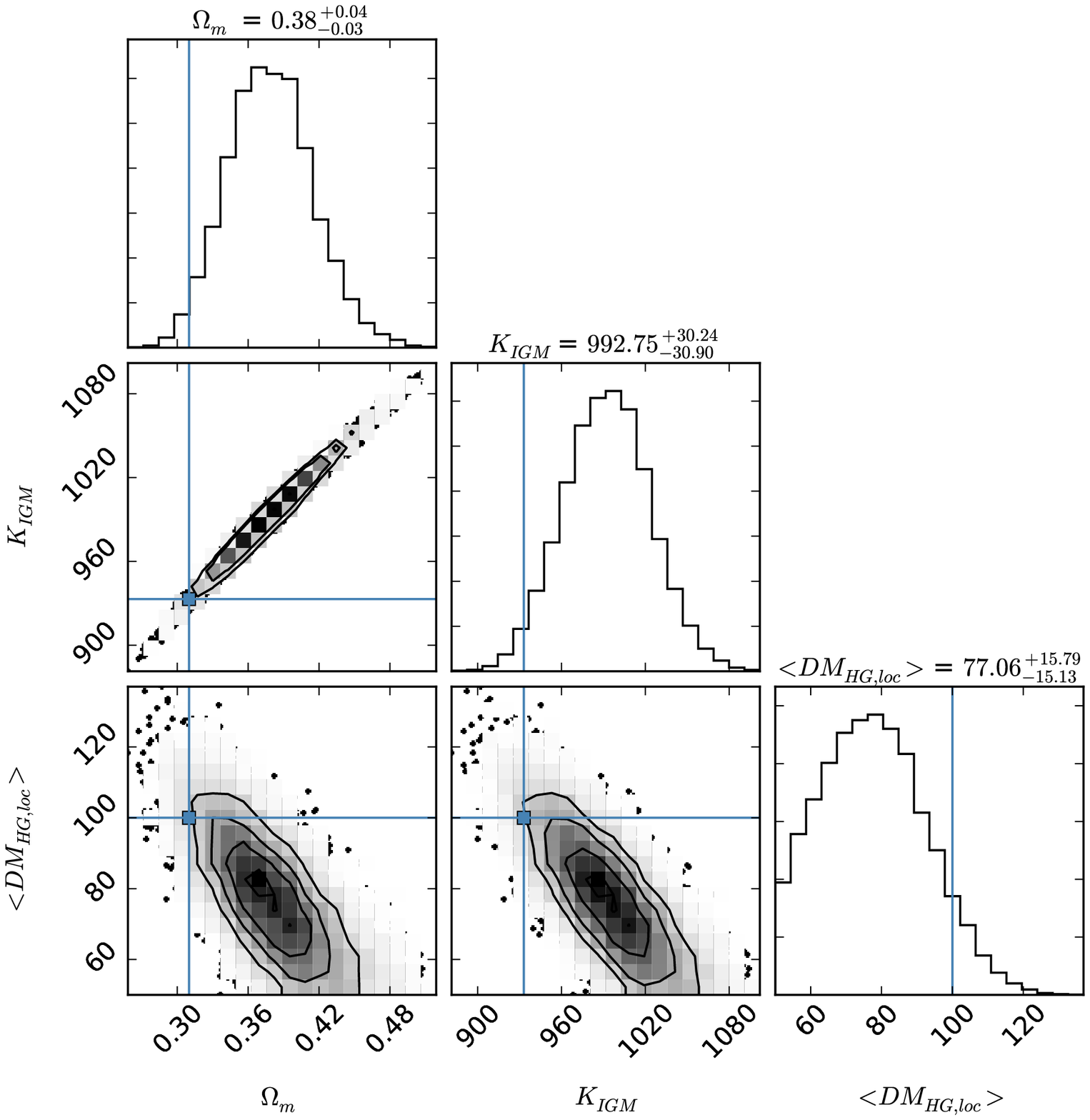}
	\end{subfigure}
	\begin{subfigure}[]
	\centering
	\includegraphics[angle=0,scale=0.3]{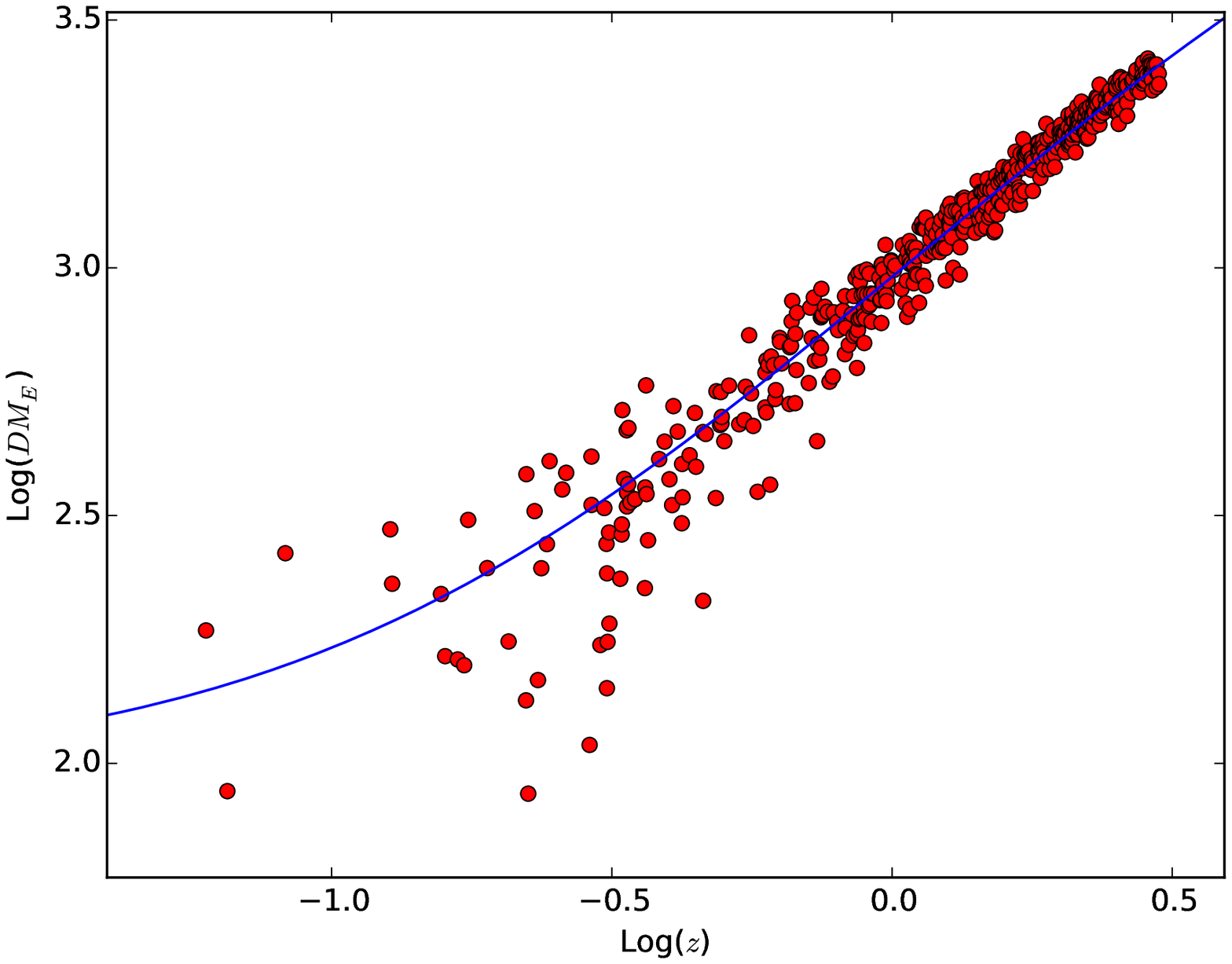}
	\end{subfigure}
	\begin{subfigure}[]
	\centering
	\includegraphics[angle=0,scale=0.3]{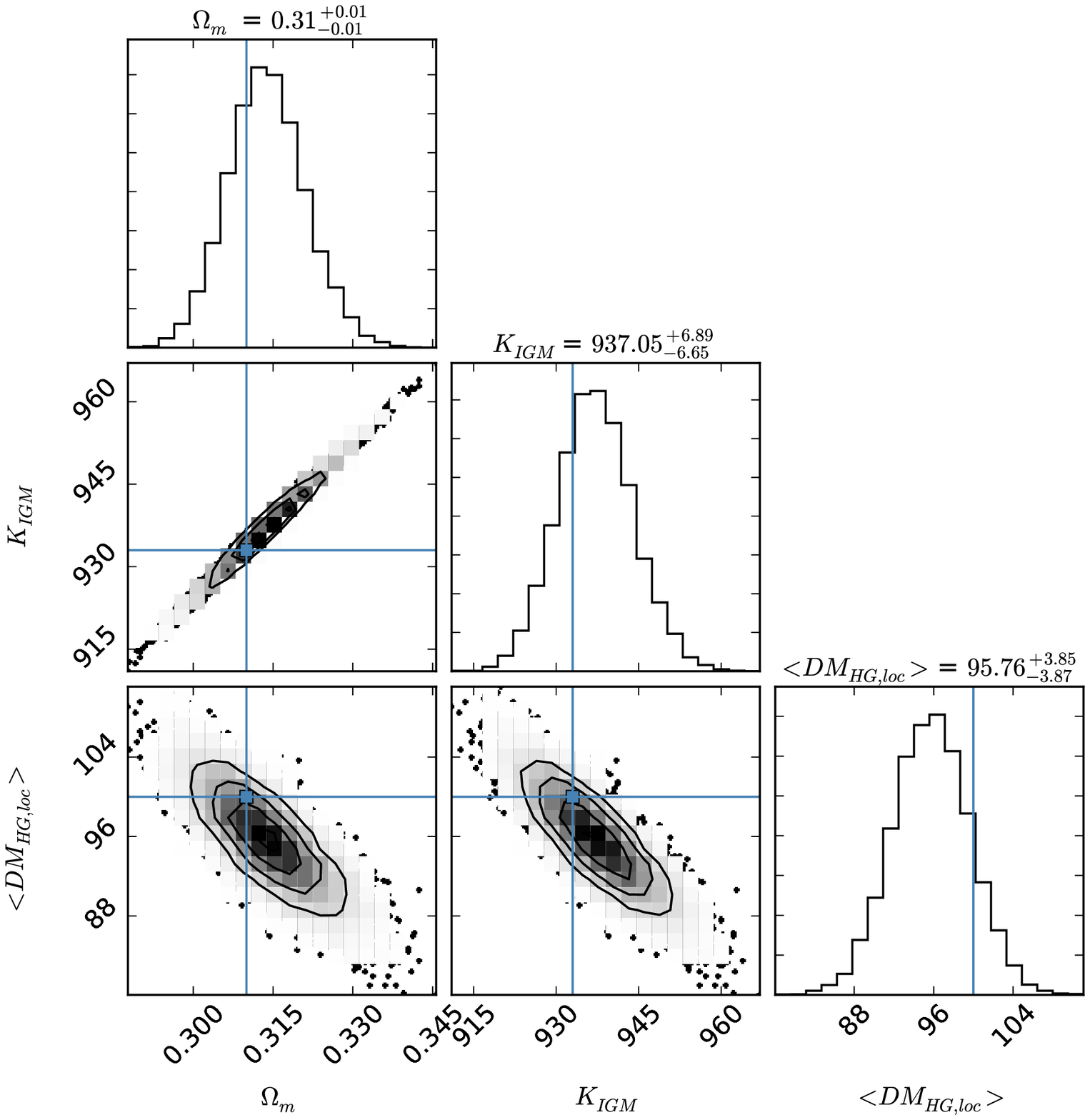}
	\end{subfigure}
\caption{Top panel: (a). The red dots denote the simulated FRB data with $N_{\rm{FRB}}=50$. The blue line denotes the MCMC best fitting curve. (b). One and two dimensional projections of the posterior probability distributions of the fitting parameters. By default, data points are shown as gray scale points with contours. Contours are shown at 0.5, 1, 1.5, and 2 $\sigma$. The blue lines denote the true values. The best fitting values are shown on top of each 1D distribution. Bottom panel: Same as top panel but for $N_{\rm{FRB}}=500$.  We assumed that $z_f=3$ and ${\rm DM_{HG,loc}}=N(100~\unit{pc~cm^{-3}},20~\unit{pc~cm^{-3}})$.}\label{fig3}
\end{figure}

To prove this, in this section we apply Monte Carlo simulations to show that one can use the MCMC method to infer the three unknown parameters. 
We adopt the flat ${\rm \Lambda CDM}$ parameters recently derived from the \emph{Plank} data: $H_0=67.7~\unit{km~s^{-1}Mpc^{-1}},~\Omega_m=0.31,~\Omega_\Lambda=0.69,~\Omega_b=0.049$ \citep{pla15}. For the fraction of baryon mass in IGM, we adopt $f_{\rm{IGM}}=0.83$ \citep{fuk98,shu12,den14}. As a result, one has $K_{{\rm IGM}}=933~\unit{pc~cm^{-3}}$. 
We assume that the redshift distribution of FRBs satisfies $P(z)=ze^{-z}$ \citep{zho14,sha11}, a phenomenological model for GRB redshift distribution. Since GRBs trace star formation history of universe, this model may stand for all FRB models invoking associations of FRBs with star formation. The true redshift distribution of FRBs depends on the underlying progenitor system(s) of FRBs, which can take different forms from this simple formula (e.g. for the $z$ distributions tracing star formation or compact star mergers, see approximate analytical expressions in \citet{sun15}). However, different models only slightly modify the distributions of $z$ of the simulated samples, but would not affect the global shape and scatter of the $\log {\rm DM_E}-\log z$ plot we are modeling. As a result, for the purpose of the simulations here, the explicit form of $z$ distribution does not affect the results. We generate a population of $N_{\rm{FRB}}$ FRBs at different redshifts between $0<z<z_f$, where $z_f$ is the redshift cutoff. At higher redshifts $z>z_f$, FRBs might be too dim to detect. Also larger DM values would make the pulses more dispersed to evade detection.
Because ${\rm DM_{MW}}$ is reasonably known, we simulate ${\rm DM_E}={\rm DM_{IGM}}+{\rm DM_{HG,loc}}/(1+z)$. We assume a normal distribution of ${\rm DM_{IGM}}=N(\langle{\rm DM}_{\rm{IGM}}\rangle,\sigma_{\rm{IGM}}$), where $\langle{\rm DM}_{\rm{IGM}}\rangle$ is given by Eq.(\ref{dmigm}) and its random fluctuation $\sigma_{\rm{IGM}}=100~\unit{pc~cm^{-3}}$ is adopted. The distribution of $\rm{DM_{HG,loc}}$ is also assumed as normal. We simulate a number of $N_{\rm{FRB}}$ FRBs, and apply the model to blindly search for input parameters. The likelihood for the fitting parameters is determined by $\chi^2$ statistics, i.e.
\begin{eqnarray}
&~&\chi^2(\Omega_m,\langle{\rm DM_{HG,loc}}\rangle,K_{{\rm IGM}})\nonumber\\
&=&\sum_{{i}}\frac{({\rm DM_{E,{\it i}}-\langle DM_E\rangle})^2}{\sigma_{\rm{IGM,{\it i}}}^2+[\sigma_{\rm{HG,loc,{\it i}}}/(1+z_{i})]^2},
\end{eqnarray}
where $i$ represents the sequence of FRB in the sample. We minimize $\chi^2$, and then convert $\chi^2$ into a probability density function. We use the software \emph{emcee}\footnote{\url{http://dan.iel.fm/emcee/current.}} to obtain the probability distribution of the fitting parameters. To test the goodness of the method, we assumed that $z_f=3$ and ${\rm DM_{HG,loc}}=N(100~\unit{pc~cm^{-3}},20~\unit{pc~cm^{-3}})$, and simulated two samples of FRBs. The first sample has $N_{\rm{FRB}}=50$ and the latter has $N_{\rm{FRB}}=500$. The analysis results are presented in the top panel of Fig.\ref{fig3} for $N_{\rm{FRB}}=50$, which give $\Omega_m=0.38^{+0.04}_{-0.03}$, $\langle\rm{DM_{HG,loc}}\rangle=77.06^{+15.79}_{-15.13}~\unit{pc~cm^{-3}}$ and $K_{{\rm IGM}}=992.75^{+30.24}_{-30.90}~\unit{pc~cm^{-3}}$. These values are all close to the initial input parameters, suggesting that the MCMC method is a powerful tool to extract the three unknown parameters. For $N_{\rm{FRB}}=500$, as shown in the bottom panel of Fig.\ref{fig3}, we obtain $\Omega_m=0.31^{+0.01}_{-0.01}$, $\langle\rm{DM_{HG,loc}}\rangle=95.76^{+3.85}_{-3.87}~\unit{pc~cm^{-3}}$ and $K_{{\rm IGM}}=937.05^{+6.89}_{-6.65}~\unit{pc~cm^{-3}}$. The results are even closer to the input values. In Fig.\ref{fig3}, the contours are shown at 0.5, 1, 1.5, and 2 $\sigma$, respectively.

\begin{figure}[H]
\centering
	\begin{subfigure}[]
	\centering
	\includegraphics[angle=0,scale=0.3]{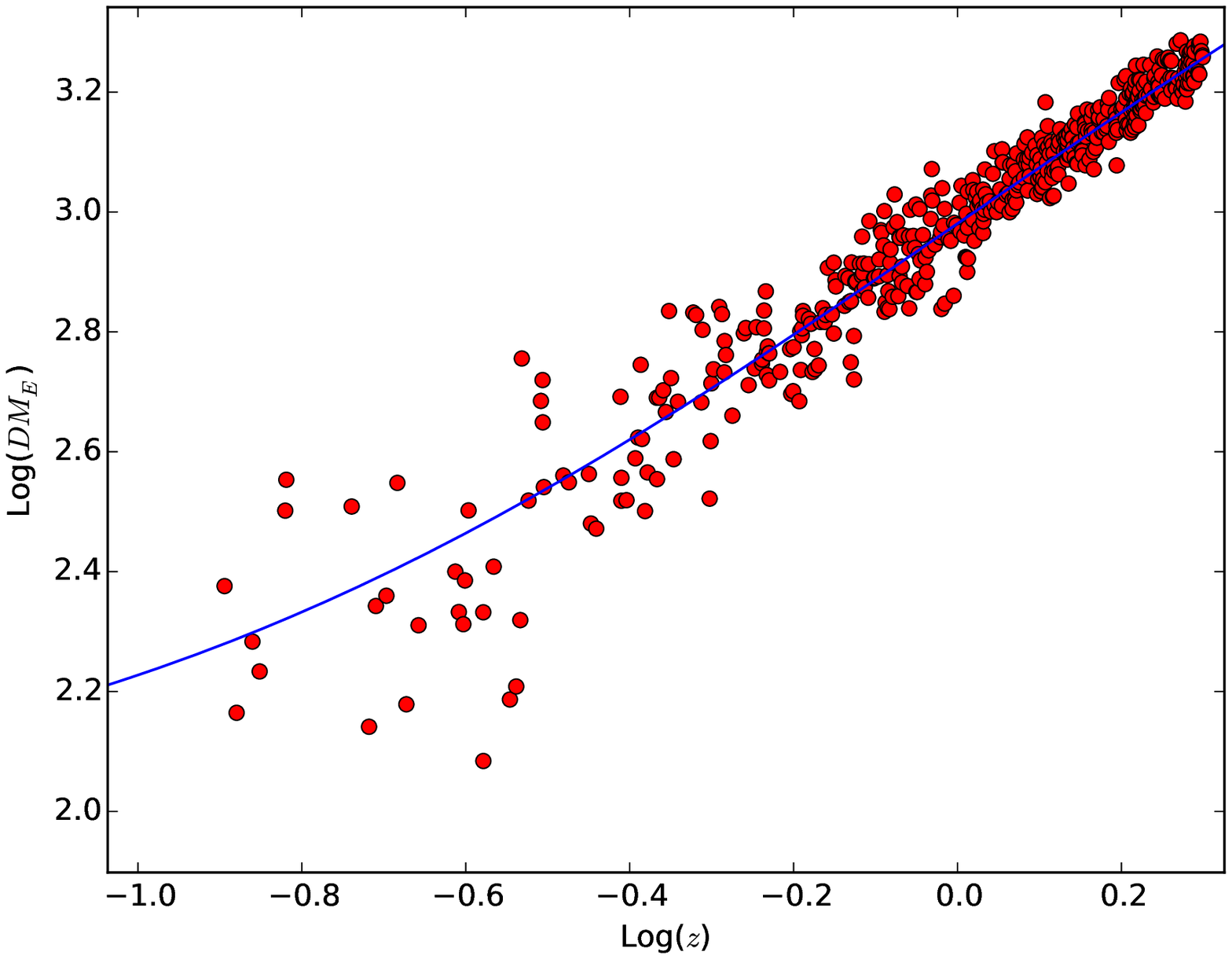}
	\end{subfigure}
	\begin{subfigure}[]
	\centering
	\includegraphics[angle=0,scale=0.3]{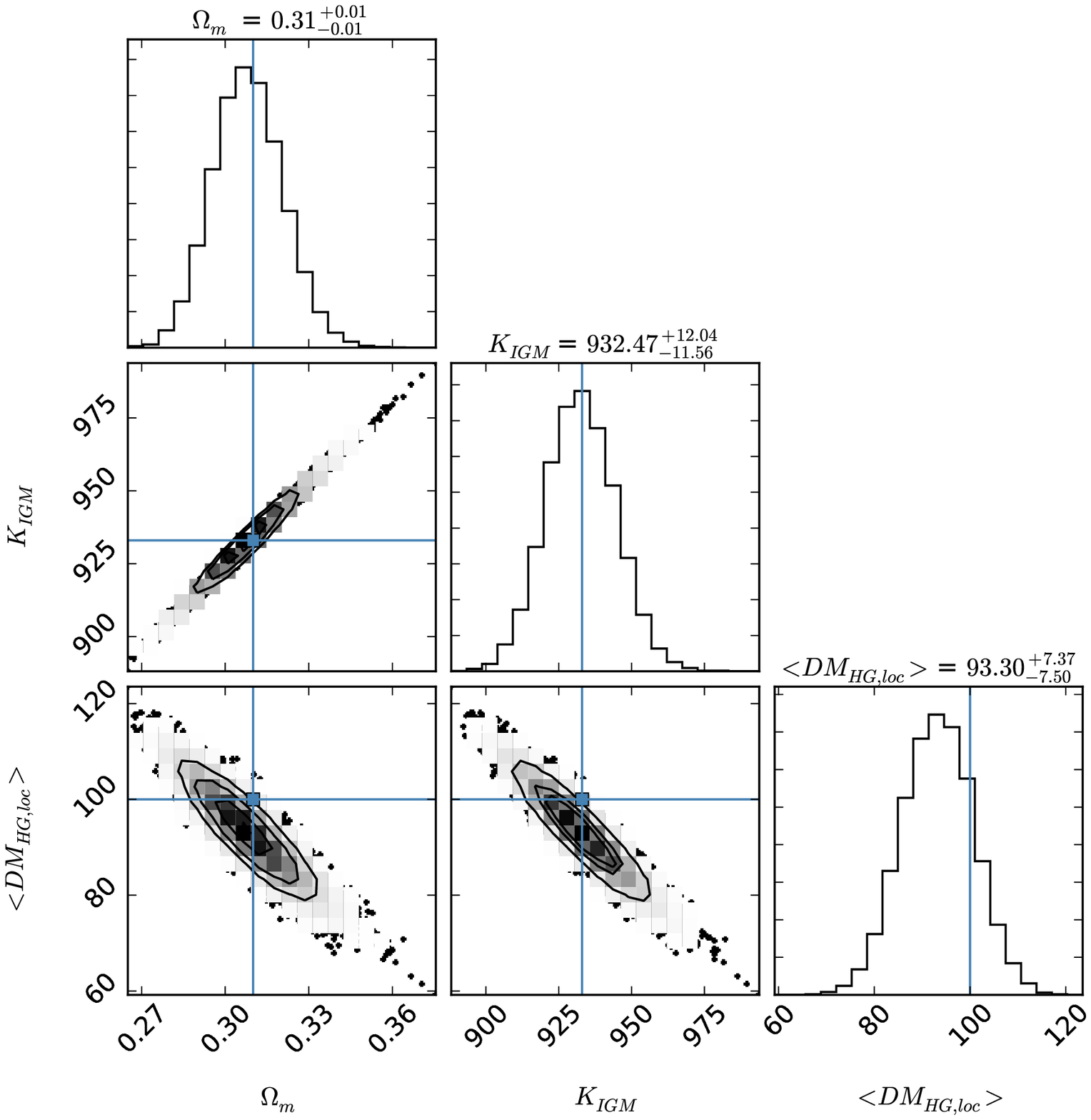}
	\end{subfigure}
	\begin{subfigure}[]
	\centering
	\includegraphics[angle=0,scale=0.3]{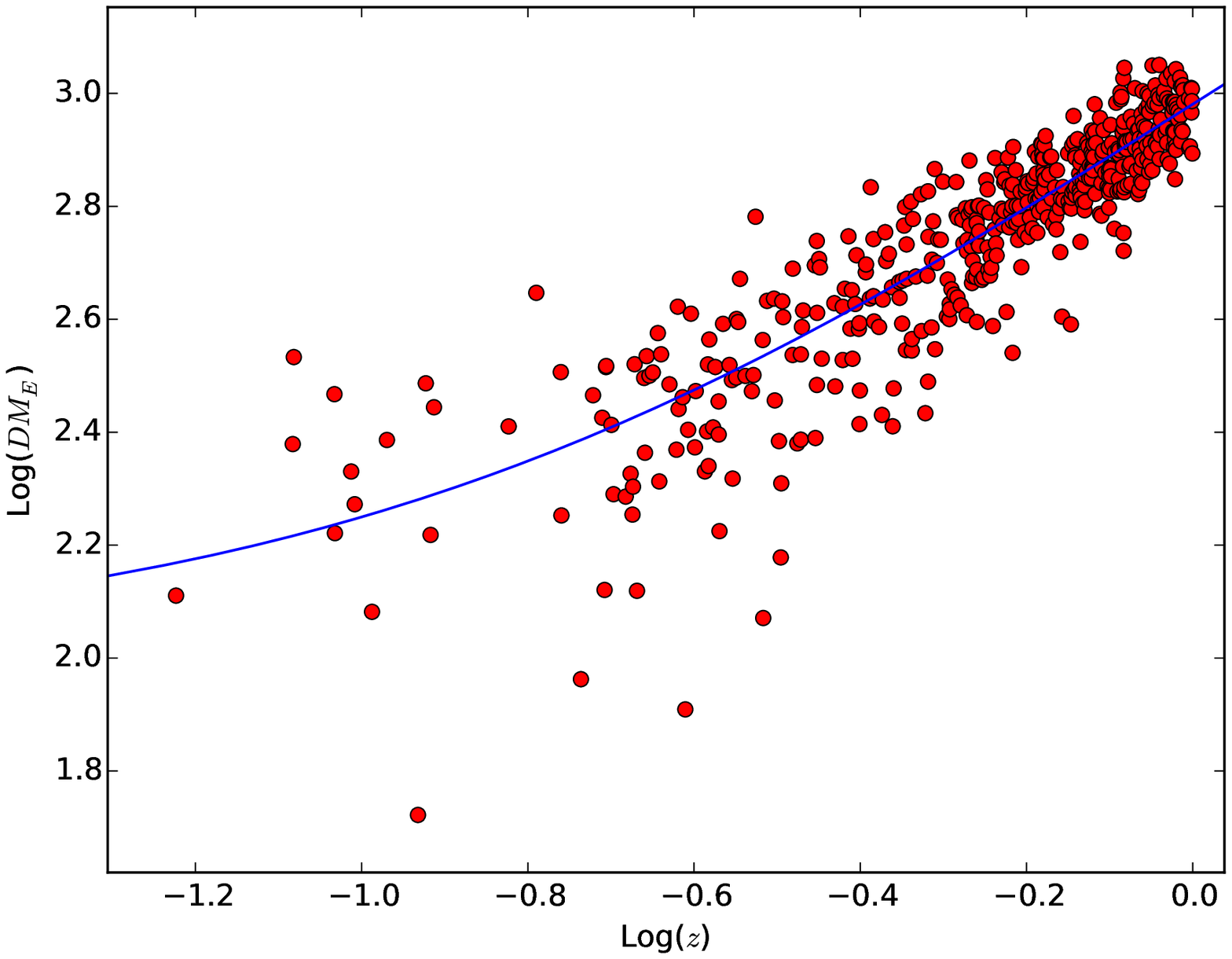}
	\end{subfigure}
	\begin{subfigure}[]
	\centering
	\includegraphics[angle=0,scale=0.3]{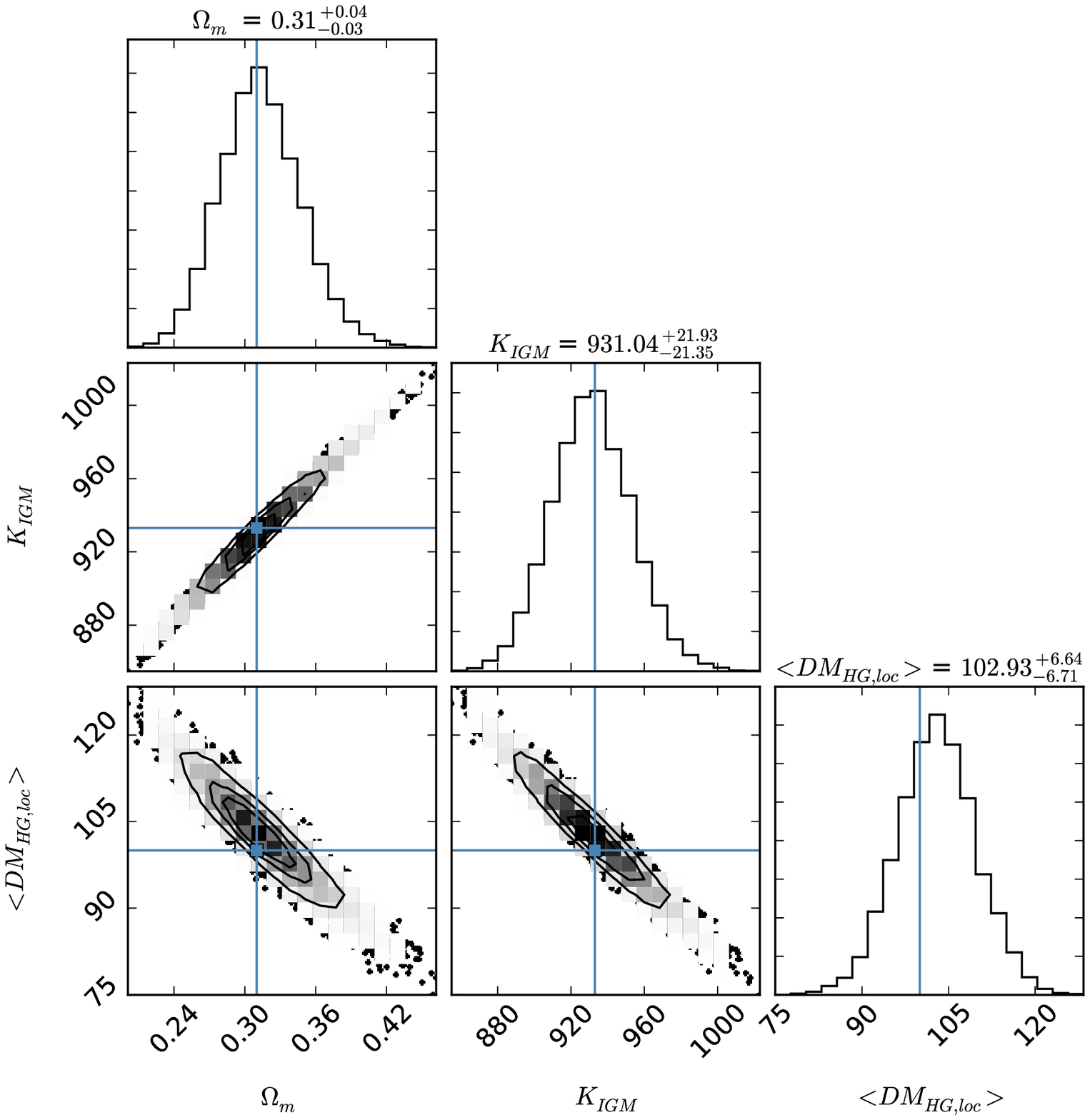}
	\end{subfigure}
\caption{Same as Fig.\ref{fig3} but for $z_f=2$ (top panel) and $z_f=1$ (bottom panel).  $N_{\rm{FRB}}=500$ and ${\rm DM_{HG,loc}}=N(100~\unit{pc~cm^{-3}},20~\unit{pc~cm^{-3}})$ are adopted.}
\label{fig4}
\end{figure}

In order to analyze the effect of $z_f$, we perform simulations with $z_f=2$ and $z_f=1$. We also assume that ${\rm DM_{HG,loc}}=N(100~\unit{pc~cm^{-3}},20~\unit{pc~cm^{-3}})$ and $N_{\rm{FRB}}=500$. 
For $z_f=2$, as shown in the top panel of Fig.\ref{fig4}, we obtain $\Omega_m=0.31^{+0.01}_{-0.01}$, $\langle\rm{DM_{HG,loc}}\rangle=93.30^{+7.37}_{-7.50}~\unit{pc~cm^{-3}}$ and $K_{{\rm IGM}}=932.47^{+12.04}_{-11.56}~\unit{pc~cm^{-3}}$.
For $z_f=1$, as shown in the bottom panel of Fig.\ref{fig4}, we obtain $\Omega_m=0.31^{+0.04}_{-0.03}$, $\langle\rm{DM_{HG,loc}}\rangle=102.93^{+6.64}_{-6.71}~\unit{pc~cm^{-3}}$ and $K_{{\rm IGM}}=931.04^{+21.93}_{-21.35}~\unit{pc~cm^{-3}}$. One can see that the results are still close to the input values, even for lower cutoff values at $z_f=2$ and $z_f=1$.

\begin{figure}[H]
\centering
	\begin{subfigure}[]
	\centering
	\includegraphics[angle=0,scale=0.3]{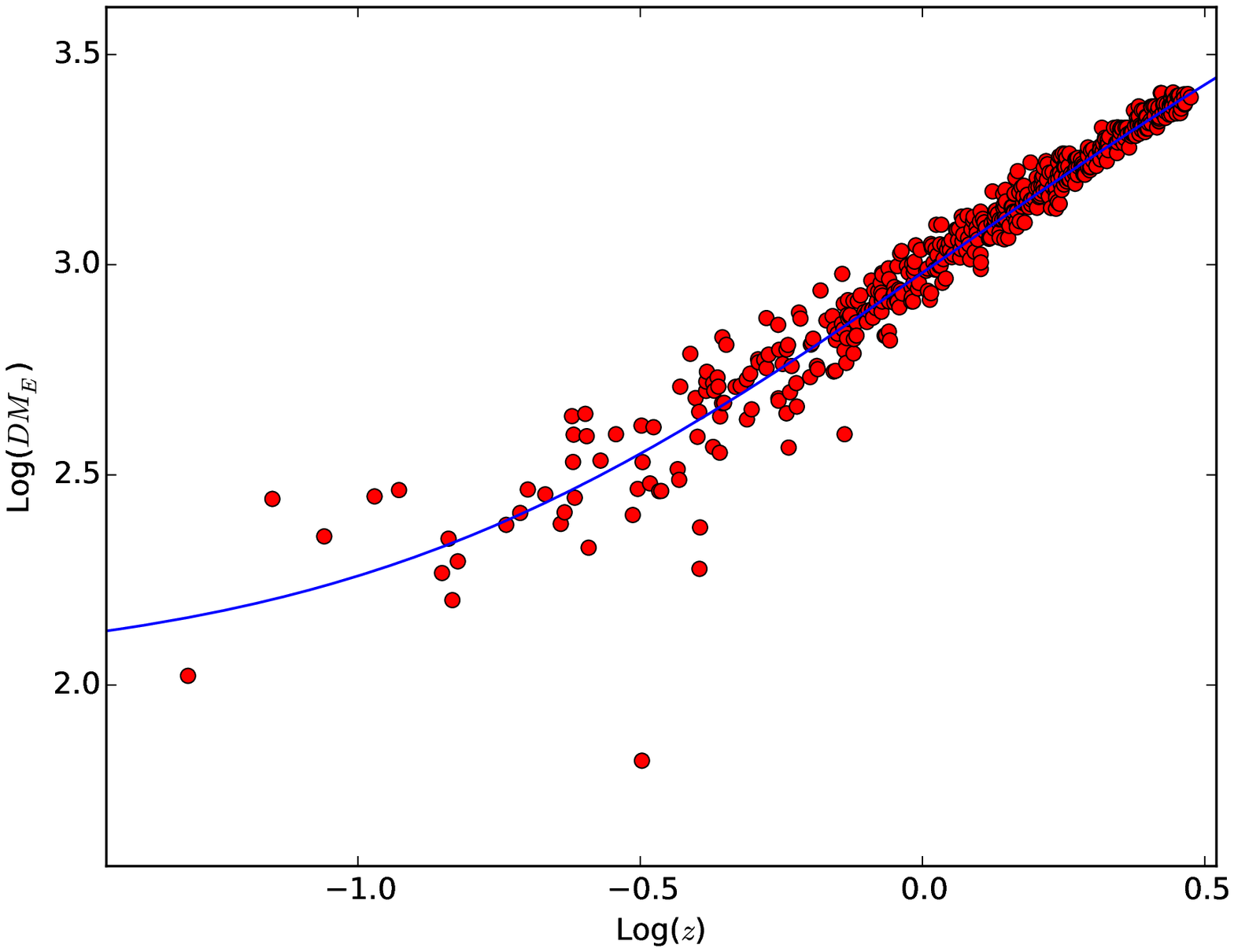}
	\end{subfigure}
	\begin{subfigure}[]
	\centering
	\includegraphics[angle=0,scale=0.3]{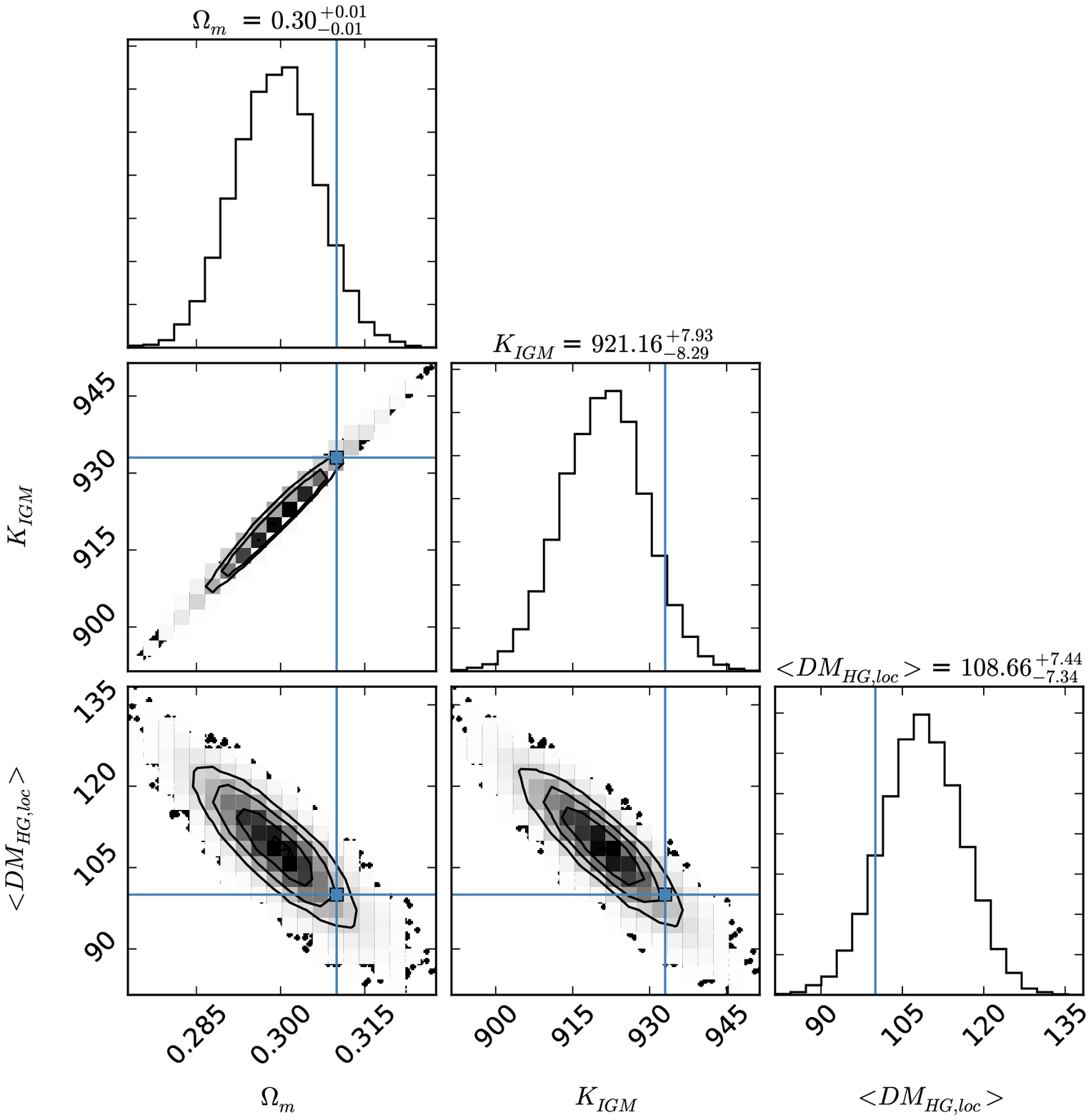}
	\end{subfigure}
	\begin{subfigure}[]
	\centering
	\includegraphics[angle=0,scale=0.3]{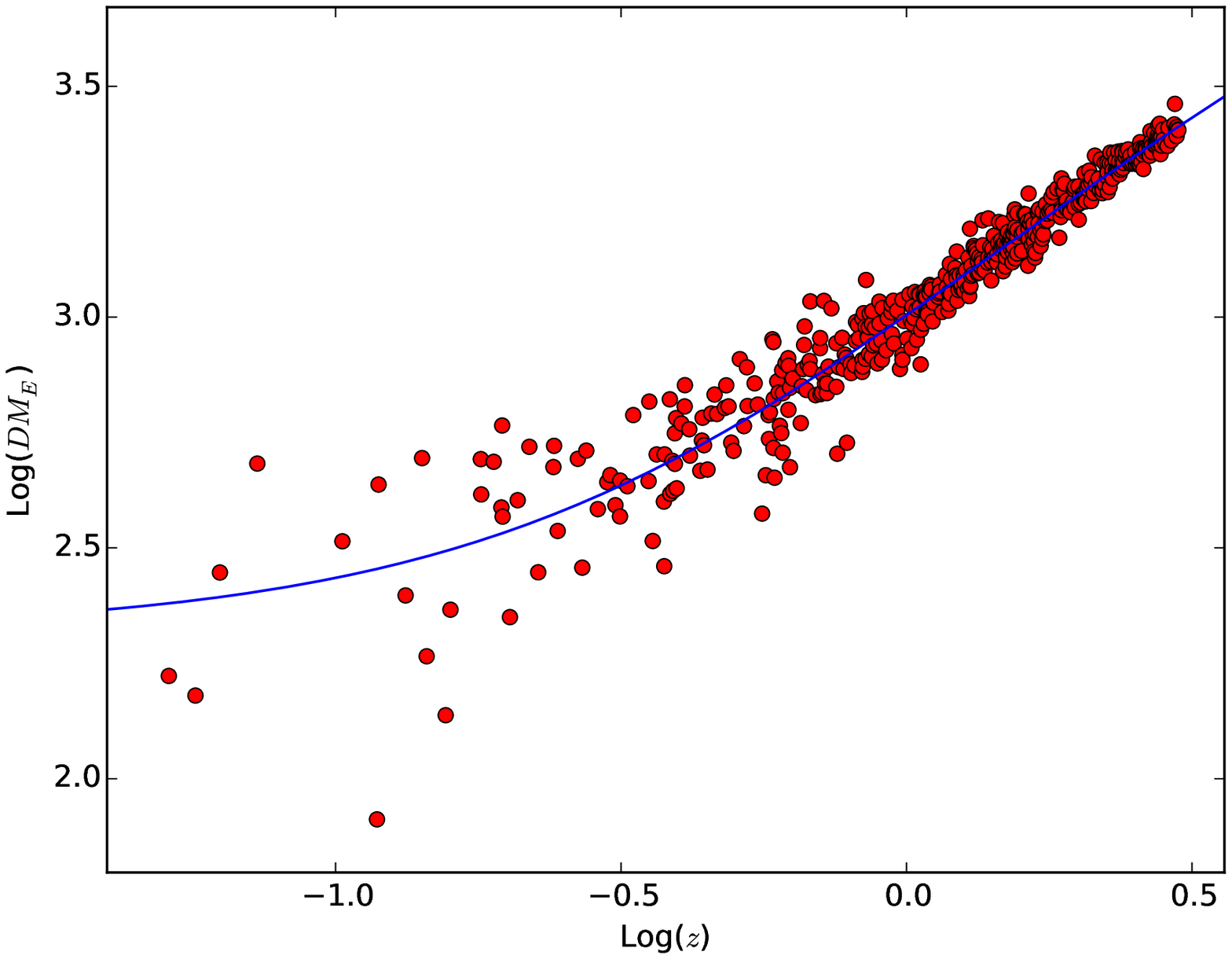}
	\end{subfigure}
	\begin{subfigure}[]
	\centering
	\includegraphics[angle=0,scale=0.3]{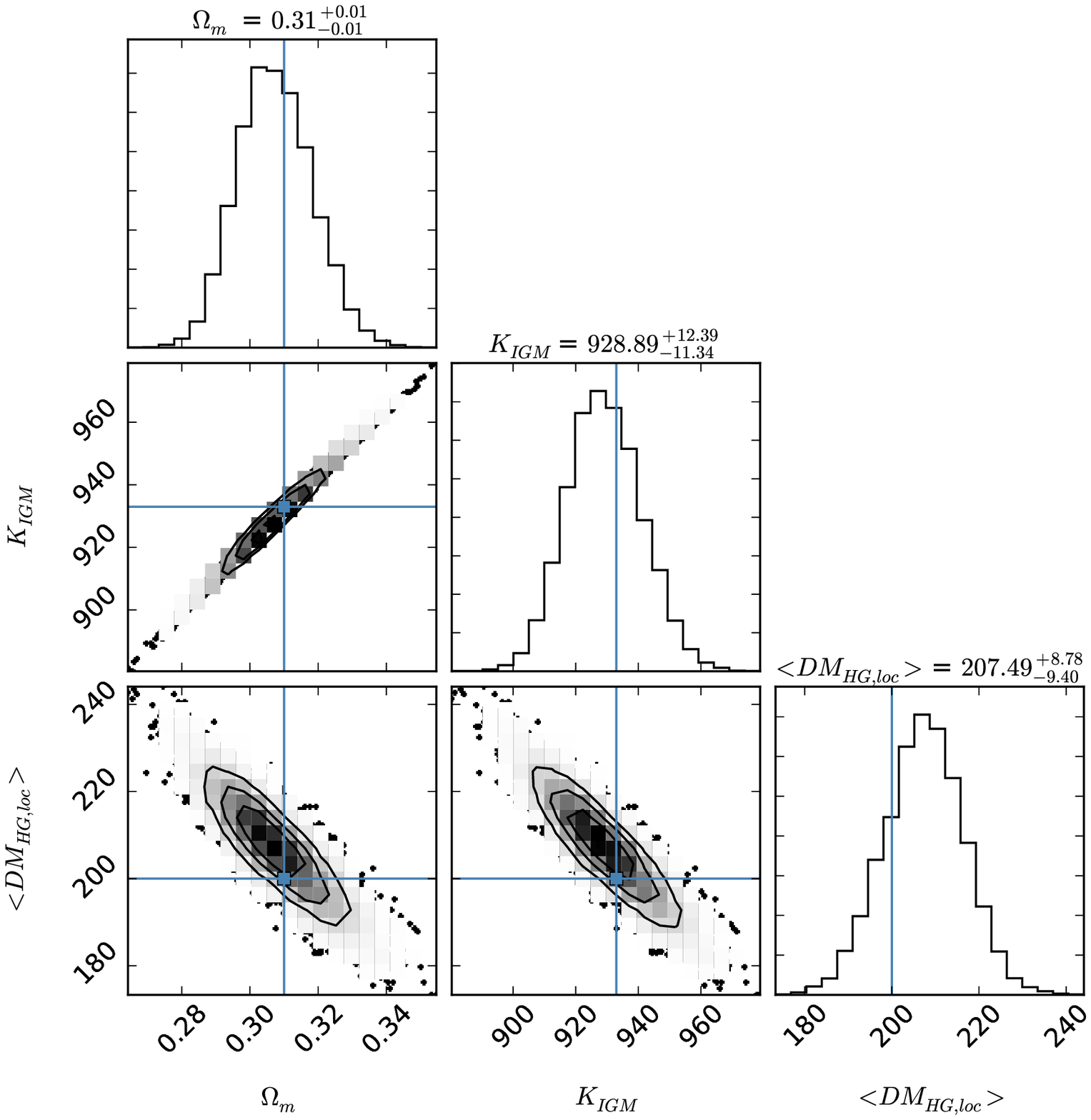}
	\end{subfigure}
\caption{Same as Fig.\ref{fig3} but for ${\rm DM_{HG,loc}}=N(100~\unit{pc~cm^{-3}},50~\unit{pc~cm^{-3}})$ (top panel) and   ${\rm DM_{HG,loc}}=N(200~\unit{pc~cm^{-3}},50~\unit{pc~cm^{-3}})$ (bottom panel). $N_{\rm{FRB}}=500$ and $z_f=3$ are adopted.}\label{fig5}
\end{figure}

Next, we test how the range of ${\rm DM_{HG,loc}}$ affects the results. We fix $z_f=3$ and $N_{\rm{FRB}}=500$, and perform simulations with ${\rm DM_{HG,loc}}=N(100~\unit{pc~cm^{-3}},50~\unit{pc~cm^{-3}})$ and ${\rm DM_{HG,loc}}=N(200~\unit{pc~cm^{-3}},50~\unit{pc~cm^{-3}})$. 
For ${\rm DM_{HG,loc}}=N(100~\unit{pc~cm^{-3}},50~\unit{pc~cm^{-3}})$, as shown in the top panel of Fig.\ref{fig5}, we obtain $\Omega_m=0.30^{+0.01}_{-0.01}$, $\langle\rm{DM_{HG,loc}}\rangle=108.66^{+7.44}_{-7.34}~\unit{pc~cm^{-3}}$ and $K_{{\rm IGM}}=921.16^{+7.93}_{-8.29}~\unit{pc~cm^{-3}}$.
For ${\rm DM_{HG,loc}}=N(200~\unit{pc~cm^{-3}},50~\unit{pc~cm^{-3}})$, as shown in the bottom panel of Fig.\ref{fig5}, we obtain $\Omega_m=0.31^{+0.01}_{-0.01}$, $\langle\rm{DM_{HG,loc}}\rangle=207.49^{+8.78}_{-9.40}~\unit{pc~cm^{-3}}$ and $K_{{\rm IGM}}=928.89^{+12.39}_{-11.34}~\unit{pc~cm^{-3}}$.
Our results show that for a certain average value, a larger random fluctuation $\sigma_{\rm{HG,loc}}$ leads to a larger systematic error of $\langle{\rm DM_{HG,loc}}\rangle$, but the inferred parameters are still close to the input values. On the other hand, for a certain $\sigma_{\rm{HG,loc}}$, the average value has little effect on the systematic error of $\langle{\rm DM_{HG,loc}}\rangle$ but does affect that of $K_{\rm{IGM}}$.

\section{Conclusion and Discussion}

In this paper, we discuss how to apply DM and $z$ information of future FRBs to study cosmology. 
Different from previous methods \citep{den14,gao14,zho14}, we do not need to assume the very uncertain host galaxy contribution to DM in the FRB sample. Instead, we show that by considering the slope parameter $\beta$, one may estimate the mean value of host DM, $ \left<{\rm DM_{HG,loc}}\right>$, using a sample of low-$z$ FRBs. Combining with FRBs detected at relatively high-$z$ ($z > 1$), one may also constrain $\Omega_m$ (within the framework of the flat $\Lambda$CDM model) and $K_{\rm IGM}$ (and hence, $\Omega_b f_{\rm IGM}$). This is because the three parameters mainly define three different properties of the ${\rm DM_E}-z$ relation: 
$\Omega_m$ defines the high-$z$ slope, $K_{\rm IGM}$ defines the global normalization ($y$-interception) of the plot in the high-$z$ regime, and ${\rm DM_{HG,loc}}$ (along with $K_{\rm IGM}$) defines the low-$z$ slope and normalization.
We perform Monte Carlo simulations to verify our claim, and find that ${\rm DM_{HG,loc}}$ and cosmological parameters can be indeed extracted from a sample of FRB using MCMC fitting.

Deriving ${\rm DM_{HG,loc}}$ from the data plays an essential role to identify the progenitor systems of FRBs. 
In our definition, ${\rm DM_{HG,loc}}$ includes the interstellar medium of the FRB host galaxy and near-source plasma. If FRB hosts are Milky-Way-like, since most FRBs come out from high latitudes from their host galaxies, the contribution from the host ISM would be much less than $100~\unit{pc~cm^{-3}}$.
If one measures $\langle{\rm DM_{HG,loc}}\rangle\gg100~\unit{pc~cm^{-3}}$ in the future, the main contribution of ${\rm DM_{HG,loc}}$ would be from the near-source plasma. The value of $\langle{\rm DM_{HG,loc}}\rangle$ would therefore place constraints on the various FRB models proposed in the literature
\citep{lor07,pop10,kea12,tho13,kas13,tot13,fal14,kul14,zha14,gen15,cor16,zha16,wan16,dai16,gu16,liu16,lyu16,kat16,mur16,pop16,pir16}.

Obtaining a reasonably large sample of FRBs with $z$ measurements may not be easy, due to the lack of a bright counterpart in other electromagnetic wavelengths hours after the burst \citep{pet15}. There are three possibilities to identify FRB redshifts in the future: 1. With very-long-baseline interferometry (VLBI) observations, one may pin down the precise location (and therefore a possible host galaxy) of an FRB, especially for dedicated observations on the repeating FRBs such as FRB 121102 \citep{spi16}; 2. Shorten the delay time of follow-up observations, and try to perform multi-wavelength follow-up observations within minutes after the FRB trigger to catch the afterglow in the brightest phase \citep{yi14}; 3. Appeal to operation of wide-field FRB search and wide-field X-ray, optical surveys to increase chance coincidence of detecting FRB counterparts during the prompt phase to catch the bright early afterglow \citep{yi14} or prompt FRB emission in other wavelengths \citep{lyu16b}. In any case, in the next few years, a few reasonable host galaxy candidates within the positional uncertainty of some FRBs may become available, so that the analysis proposed in this {\em Letter} may be carried out.

\acknowledgments
We thank the anonymous referee for detailed suggestions that have allowed us to improve this manuscript significantly. We also thank Zhuo Li, Hai Yu and Bin-Bin Zhang for helpful comments and discussion. This work is partially supported by The Initiative Postdocs Supporting Program and the National Basic Research Program (973 Program) of China (Grant 2014CB845800).

\end{document}